# Resolving the Quantum Measurement Problem through Leveraging the Uncertainty Principle

Kyoung Yeon Kim

Department of Electrical and Computer Engineering, Seoul National University, South Korea

**Abstract**

The Schrödinger equation is incomplete, inherently unable to explain the collapse of the wavefunction caused by measurement—a fundamental issue known as the "quantum measurement problem." Quantum mechanics is generally constrained by the uncertainty principle and, therefore, cannot interpret definite observations without uncertainty. Here, we resolve this enigma by demonstrating that in phase-space quantum mechanics, particularly through the Wigner-Moyal equation, uncertainty can be arbitrarily adjusted by tuning the "observation window." An observation window much smaller than the uncertainty limit causes substantial nonlocality, rendering the problem ill-posed. This suggests that only with sufficient uncertainty does nonlocality become bounded, resulting in a well-posed universe. Conversely, in the absence of uncertainty, spacetime is warped beyond recognition, and the system exists as a superposition of numerous possible states. Measurement collapses this superposition into a unique solution, exhibiting timeless nonlocal interactions. Our framework bridges seemingly disparate concepts such as classical mechanics, quantum mechanics, decoherence, and measurement within intrinsic quantum mechanics even without invoking new theory.

**Main**

Wave-particle duality is famously demonstrated in the double-slit experiment, where electrons exhibit wave-like behavior, forming an interference pattern[1,2]. However, when an electron passing through the slits is observed, this pattern disappears, and the electron exhibits particle-like properties[3,4]. According to the Copenhagen interpretation[5,6], the electron exists as a probability distribution until it is observed, at which point the wavefunction collapses, and the electron assumes a specific location. Ironically, well-known quantum mechanical equations, such as the Schrödinger equation[7], describe the wave nature of electrons but offer no explanation for what happens during the measurement process. For instance, in flipping a coin, quantum mechanics provides a single solution predicting a 50% probability for heads and tails, yet the actual measurement yields one of two distinct outcomes—heads or tails. This disparity gives rise to the "quantum measurement problem," an unresolved and often philosophical question that remains under debate[8,9]. Quantum decoherence, an alternative approach, posits that the quantum-to-classical transition results from interactions between the system and its environment, causing the loss of quantum coherence[10,11]. While insightful, this framework does not explain the actual collapse of the wavefunction and is therefore not generally regarded as a resolution to the quantum measurement problem[12,13]. Can all these issues be addressed within the framework of intrinsic quantum mechanics, without invoking speculative new formulations?

Here, we demonstrate that the quantum measurement problem can be resolved using a single equation in phase-space quantum mechanics, without introducing any additional terms. In 1932, Eugene Paul Wigner formulated the quantum mechanical distribution function in phase space as a Fourier transform of the density matrix[14]:

$$f(x, k, t) = \int_{-\infty}^{\infty} e^{-iky} \rho\left(x + \frac{y}{2}, x - \frac{y}{2}, t\right) dy. \quad (1)$$

This distribution takes the same form as a classical probability distribution but, due to strong quantum effects, can locally assume negative values, earning it the name "quasi-probability distribution." The propagation of the Wigner distribution function can be expressed as the Moyal equation[15]:

$$\frac{\partial f(x,k,t)}{\partial t} + \frac{\hbar k}{m}\nabla_x f(x,k,t) = \sum_{j=0}^{\infty}\frac{(-1)^j}{4^j(2j+1)!\,\hbar}\frac{\partial^{2j+1}U(x,t)}{\partial x^{2j+1}}\frac{\partial^{2j+1}f(x,k,t)}{\partial k^{2j+1}}. \quad (2)$$

Thanks to its simplicity and resemblance to the classical Liouville equation, the Moyal equation has gained prominence and been widely used, particularly in deriving quantum-corrected models[16,17]. Ironically, however, direct numerical solutions of this equation have been notoriously difficult for over 70 years. In cases involving singular potential profiles, whose derivatives survive up to very high orders, solving equation (2) is known to be both troublesome and unstable[18].

Unlike the Schrödinger equation, the process of solving the Wigner-Moyal equation does not inherently incorporate the constraints of the uncertainty principle[19,20], rendering stable simulations impossible. To address this, we developed a completely novel approach to freely tune uncertainty using the new concept of an "observation window." This approach not only enables stable simulations but also reveals intriguing physical characteristics. Fig. 1 illustrates our key findings. In macroscopic systems, nonlocality vanishes, and the system adheres to classical dynamics. In microscopic systems, however, a sufficiently wide observation window within the uncertainty limit results in coherent behavior, whereas a narrow observation window (low uncertainty) leads to quantum decoherence. Moreover, as the observation window approaches zero and uncertainty disappears, the solution aligns with the measurement problem. This demonstrates that classical mechanics, quantum mechanics, decoherence, and the measurement problem can all be unified and interpreted through a single intrinsic quantum mechanical equation, the Wigner-Moyal equation. In the following sections, we provide a detailed explanation of this discovery.

**Failure of Microscopic Numerical Observation**

We begin by introducing the general discretization scheme and the resulting numerical challenges. Numerical observation was performed by solving equation (2) using the finite difference method (FDM) with mesh resolutions of $\Delta x$ and $\Delta k$. In the steady state, equation (2) is discretized with the lowest-order accuracy, a standard approach for solving partial differential equations (PDEs):

$$\frac{\hbar k}{m} \frac{\mp f(x,k) \pm f(x \pm \Delta x, k)}{\Delta x} = \sum_{j=0}^{\infty} \left[ \frac{(-1)^j}{\hbar \Delta x \Delta k} C_{2j+1} U_{2j+1} \sum_{l=-2j-2}^{2j+2} a_{2j+1,2}^l f(x, k + l\Delta k) \right]. \quad (3)$$

$$C_j = \frac{1}{j! (2\Delta x \Delta k)^{j-1}}. \quad (4)$$

Here, $a_{j,m}^l$ represents the finite difference coefficients for the $j$-th order and $m$-th accuracy, $U_j$ is the numerical gradient of the potential energy, and $C_j$ denotes the nonlocal power. The upper and lower signs on the left-hand side (LHS) correspond to upwind and downwind schemes, respectively. This equation solves for the unknown $f(x,k)$, with the remaining terms being explicitly defined. This represents a "numerical observation" of the distribution function spread across k-space at a specific position within the range defined by $\Delta x$.

Fig. 2a shows the nonlocal power $C_j$ for various $\Delta x \Delta k$ values. As $\Delta x \Delta k$ increases, the zero-order terms become dominant while higher-order terms diminish. In such case, $2\Delta x \Delta k$ in equation (4) exceeds 1, causing $C_j$ to decay rapidly. Under these conditions, equation (3) can be approximated as:

$$\frac{\hbar k}{m} \frac{\mp f(x,k) \pm f(x \pm \Delta x, k)}{\Delta x} \approx \frac{U_1}{\hbar \Delta x \Delta k} \sum_{l=-1}^{1} a_{1,2}^l f(x, k + l\Delta k). \quad (5)$$

This is equivalent to the discretized classical Boltzmann transport equation[21]. Thus, it can be demonstrated that quantum mechanics and classical physics converge at sufficiently large mesh spacing. Similar to how quantum effects are generally unobservable in macroscopic systems, quantum mechanics lack significant nonlocal terms on macroscopic scales, resulting in behavior indistinguishable from classical physics, as illustrated in Fig. 1.

Fig. 2b shows that when $\Delta x \Delta k$ becomes much smaller than 0.5, the distribution of $C_j$ broadens significantly, and its peak value increases rapidly. This indicates that as the mesh size decreases, the system matrix becomes more complex due to enhanced nonlocality, requiring interactions with a greater number of nodes. A fortunate aspect of equation (4) is that, regardless of how small the mesh size becomes, its form remains similar to a Poisson distribution, allowing it to be reduced to a finite-order PDE by truncating derivative terms based on a cutoff value. However, a significant issue arises when $\Delta x \Delta k$ becomes too small: the peak value of $C_j$ grows excessively, reaching unphysical levels. In such cases, $C_j$ becomes much larger than $C_0$ and the nonlocal effects overwhelm the classical kinetic terms.

To investigate numerical behavior under these conditions, we conducted simulations for various mesh sizes. Using a simplified resonant tunneling diode with a linear potential drop in the double barrier region (Fig. 2c), we evaluated quantum effects by fixing $\Delta x$ at $0.4\ nm$ and varied $\Delta k$. As anticipated, as $\Delta x \Delta k$ decreases, the simulation results become increasingly unphysical, ultimately leading to a singularity occurs the system matrix. Specifically, as the mesh size shrinks, negative electron density appears (Fig. 2c), and the resonance effect diminishes, eventually producing oscillatory and unstable results, as shown in Fig. 2d. Extreme nonlocality renders numerical solutions to equation (4) unstable at sufficiently small mesh sizes. This behavior is abnormal and contrary to conventional PDEs, where accuracy and stability typically improve as the mesh size decreases[22]. Since equation (4) is clearly physical, we hypothesized that there must be a workaround. So, what is causing this troublesome numerical issue?

**Numerical Uncertainty Principle**

We postulate that nonphysical results arise because using excessively small mesh sizes violates the uncertainty principle. When computing each differential term in equation (3), nonlocal data points within a specific range are utilized, as illustrated in Fig. 3a. We define this range as the "observation window." Our findings indicate that when the observation window becomes narrower than the uncertainty limit, substantial nonlocality emerges, leading to ill-posed scenarios. The uncertainty principle states that it is impossible to predict the exact position and momentum of a particle, and instead describes the system as a superposition of multiple available states represented by a probability function that reflects their correlations. However, using low-order accuracy schemes that calculate derivatives by sampling data over regions much narrower than the uncertainty limit violates this principle.

To address this, we propose a new approach: employing very high-order accuracy schemes[23] that calculate differential terms using data from a much wider range, thereby effectively expanding the observation window while maintaining a small mesh size. By imposing a minimum limit $N_{Obs}$ on the observation window of the differential terms on the RHS of equation (3), we expand the narrow observation window to the desired range:

$$\sum_{j=0}^{\infty}\left[\frac{(-1)^j}{\hbar \Delta x \Delta k}C_{2j+1}U_{2j+1}\sum_{l=-N_{min}}^{N_{min}}a_{2j+1,2N_{Obs}-2j}^l f(x,k+l\Delta k)\right], for\ (N_{Obs}-j)>0. \quad (6)$$

This can be understood by considering that, as shown in Fig. 3a-b, we expand the observation windows of specific derivative terms; originally, these terms have observation windows smaller than $[-N_{min}, N_{min}]$ in the second-order accuracy scheme, which are expanded to $[-N_{min}, N_{min}]$ using the high-order scheme.

Expanding the observation window requires employing very high-order accuracy schemes, and our findings reveal a unique numerical property in such schemes. The sum of the absolute values of the finite difference coefficients for the $j$-th derivative with $m$-th order accuracy is given by:

$$A_{j,m} = \sum_{l=-j-\frac{m}{2}}^{j+\frac{m}{2}} |a_{j,m}^l| \quad (7)$$

When plotting $A_{j,m}$, we observe that in high-order schemes, $A_{j,m}$ exhibits a sharply decaying behavior as $m$ increases, as shown in SM. Fig. 1. This implies that excessive nonlocality can be effectively suppressed as the observation window expands, due to the diminishing weight of nonlocality represented by $A_{j,m}$. While it is generally understood that high-order schemes enhance numerical accuracy, our results demonstrate that for very large $j$ and $m$, their primary effect is to suppress nonlocality rather than improve accuracy. This finding shows that tuning the observation window is fundamentally an act of controlling the power of nonlocality. Interestingly, these numerical properties align with a previous study suggesting that, across all physical theories, nonlocality will decrease as uncertainty increases[24].

Since the size of the observation window determines the constraints imposed by the uncertainty principle in the Wigner-Moyal equation, we conducted simulations for various $N_{Obs}$ values. Fig. 3c shows that the most pronounced quantum coherence occurs near $N_{Obs} \approx 1/\Delta x \Delta k$. When $N_{Obs}$ exceeds this value, the uncertainty becomes too large, causing the resonance effect to fade and eventually disappear. Conversely, when $N_{Obs}$ falls below this value, the uncertainty drops below the critical limit, leading to diminished resonance effects and ultimately resulting in unphysical outcomes. Thus, we define the condition under which coherence is maximized as the numerical uncertainty principle:

$$N_{Obs} = N_{uncertainty} = \frac{1}{\Delta x \Delta k} \qquad (8)$$

This is a universal condition that can be applied to any mesh spacing. To validate its applicability, we conducted further simulations under this condition for various mesh sizes.

Fig. 4d illustrates the simulation results for various mesh sizes under the uncertainty limit ($N_{Obs} = N_{uncertainty}$). Although the Wigner-Moyal equation is known for its extreme numerical instability, as shown in Fig. 1d and Fig. 3c, our novel numerical scheme remarkably restores simulation consistency and stability. Because our method yields highly consistent simulation results that are largely unaffected by mesh spacing, coarse meshes can be utilized to enhance computational efficiency. This represents the first successful demonstration of the Wigner-Moyal equation as a stable quantum transport solver. This approach is particularly effective because the system matrix retains complexity only near quantum regions, while naturally reducing to sparse classical dynamics in classical regions as high-order terms vanish. SM. Fig. 2 demonstrates that by expanding $N_{Obs}$, it effectively enhances the sparsity of the system matrix by eliminating unphysical nonlocality branches that violate the uncertainty principle. Consequently, this approach can be efficiently applied to large-scale quantum-classical mixed systems of arbitrary size[25,26].

**Quantum Decoherence**

When both the mesh size (Fig. 2b) and observation window (Fig. 3c) decrease, nonlocality is amplified, diverging from classical equations and seemingly enhancing quantum mechanical effects. However, we observed that quantum effects actually become weaker in both cases. We postulate that narrowing the observation window represents a transition from a high-uncertainty probability distribution to a low-uncertainty measurement system. This offers a new perspective: that low uncertainty, resulting from a narrow observation window, fundamentally signifies quantum decoherence.

To validate this interpretation, we designed a numerical experiment inspired by the double-slit experiment, in which observing electrons passing through slits causes the interference pattern to vanish. Using a resonant tunneling diode, as shown in Fig. 4a, we simulated the effects of quantum decoherence at specific points. Remarkably, similar to the double-slit experiment, when decoherence occurs at point A, the entrance of the double barrier, the resonance characteristics are destroyed, as shown in Fig. 4b. In contrast, when decoherence occurs at point B, located after the resonance region, the resonance phenomena remain mostly intact. These findings reveal an intriguing result: decoherence at even a single mesh point can render quantum effects unobservable, akin to the double-slit experiment. In the coherent case, strong quantum interference during resonance leads to the emergence of negative values in the Wigner function, as shown in Fig. 4c. These negative values arise due to our uncertainty about the system, essentially reflecting a quasi-probability rather than an actual negative probability or density. In a sense, they can be considered illusions, since truly negative probability or density cannot exist in reality. To verify whether such negative densities are genuinely observable, we reduce the uncertainty to examine the system more closely. As shown in Fig. 4d, the negative values are suppressed, and these phenomena become unobservable. In other words, when we attempt to observe the system, quantum interference and resonance—characterized by negative quasi-probability distributions—disappear because such effects fundamentally cannot exist. This clearly demonstrates that low uncertainty directly correlates with quantum decoherence and observation.

Traditionally, quantum decoherence is understood as a quantum-to-classical transition driven by interactions with the environment. However, our results suggest that decoherence represents convergence to measurement results, characterized by reduced uncertainty and amplified nonlocality. Specifically, the observation window in this case satisfies the following condition, which violates the numerical uncertainty principle:

$$N_{Obs} < N_{uncertainty} = \frac{1}{\Delta x \Delta k}. \qquad (9)$$

This methodology introduces a powerful numerical approach that naturally tunes decoherence by simply modifying the numerical scheme used to solve the Wigner-Moyal equation, without requiring artificial additional terms. All external influences, such as scattering, inherently induce decoherence[27]. This implies that narrowing the observation window can act as an intrinsic scattering model, even in the absence of explicit interaction terms. From this perspective, reducing uncertainty is critical in most real-world scenarios involving scattering-induced decoherence. This further suggests that Schrödinger equation-based quantum dynamic models, such as non-equilibrium Green's function method[28,29], which lack the ability to tune uncertainty, may be fundamentally unsuited for addressing real-world quantum phenomena.

**Measurement without Uncertainty**

According to our interpretation, as the observation window (uncertainty) approaches zero, the system equation transitions into a perfect "measurement problem." In our previous results, we showed that when the observation window becomes too small, the simulation becomes ill-posed and unstable. Notably, this is exactly the result we expected. Quantum mechanics provides a unique solution, but fundamentally, a measurement without uncertainty is an ill-posed problem that can have numerous solutions (no unique solution). However, using a special approach, we can interpret properties of this scenario. Dividing the discrete equation using the forward Euler method by the maximum value of $C_j$, denoted as $C_{max}$, yields the following form:

$$\frac{1}{C_{max}} \left[ \frac{f(x,k,t+\Delta t_p) - f(x,k,t)}{\Delta t_p} + \frac{\hbar k}{m} \frac{\mp f(x,k,t) \pm f(x \pm \Delta x, k, t)}{\Delta x} \right]$$
$$= \sum_{j=0}^{\infty} \left[ \frac{(-1)^j}{\hbar \Delta x \Delta k} \frac{C_{2j+1} U_{2j+1}}{C_{max}} \sum_{l=-j-1}^{j+1} a_{2j+1,2}^l f(x, k+l\Delta k, t) \right], \quad (10)$$

where $\Delta t_p$ is the Planck time, the smallest physical timescale. As $\Delta x \Delta k$ decreases, $C_{max}$ approaches infinity, rendering the drift and time evolution term on the LHS negligible:

$$\sum_{j=0}^{\infty} \left[ \frac{(-1)^j}{\hbar \Delta x \Delta k} \frac{C_{2j+1} U_{2j+1}}{C_{max}} \sum_{l=-j-1}^{j+1} a_{2j+1,2}^l f(x, k+l\Delta k, t) \right] \approx 0 \quad (11)$$

This implies that the equation no longer describes quantum dynamics but instead represents an entanglement (constraint) system where all correlated information is instantaneously determined, independent of time. However, this PDE is ill-posed, lacking a unique solution. From a numerical perspective, it can be thought of as a PDE without proper boundary conditions. Unless boundary conditions are imposed at each position, the system remains in a superposition state with infinitely many solutions. If a value at a specific point is fixed (a boundary condition in numerical terms), the superposition collapses, and the solution in the remaining entangled space is immediately determined.

Equation (11) depends solely on the potential derivative $U_j$, meaning its solution could vary with different potential profiles. To test this, we generated ten random potential profiles (Fig. 5a) and performed measurements (solving Equation (9)) at position $x = 70nm$ with a Dirichlet boundary condition applied at $k = 0$, representing an observation at this momentum. As shown in Fig. 5b, the measurement results were nearly identical regardless of the potential profile, consistently producing a pronounced peak at the observation point. This indicates that when the system is perturbed by the act of measurement, the superposition that existed probabilistically under uncertainty collapses instantaneously. The state at that precise point is observed, and simultaneously, the information of all entangled regions becomes determined. Additionally, applying multiple Dirichlet boundary conditions at different points allowed the simultaneous measurement of particles with various momenta, as shown in Fig. 5c.

As the momentum mesh size decreases, the range of nonlocal interactions extends, theoretically enabling infinite nonlocal interaction as the mesh size approaches zero. The only scenario where measurement becomes impossible occurs when the potential energy derivative term vanishes, invalidating equation (9). To investigate this, we conducted a toy "Big Bang" experiment, illustrated in Fig. 5d. We simulated a null space with uniform potential energy and introduced a trigger (a potential pulse) at a distant point, equivalent to the distance from Andromeda to Earth. In the null space, as shown in Fig. 5e, particles cannot be measured. However, once the trigger occurs, particles can, in principle, be measured and generated instantaneously, even at vast distances, forming an infinite universe. In this way, while probability distributions in the real world propagate gradually over time, from a numerical perspective, measurement is a completely random act transcending spacetime, involving infinitely many possibilities. In other words, what we perceive as time is the sequential arrangement of random pictures following a certain set of rules.

**Transition from Quasi-probability to Probability by Measurement**

The solution of quantum mechanical equation can be misconsidered to represent the probability of measurement outcome. However, from the perspective of phase-space quantum mechanics, which mimics the actual measurement of position and momentum, this cannot hold true. As shown in Fig. 4c the Wigner function takes on negative value in regions of strong quantum interference, however, negative quasi-probability cannot exist in reality. Therefore, it is more reasonable to explain that classical distribution determines the probability of measurement where quantum interference vanishes. We explain this based on our previous approaches:

$$\frac{f(x,k,t+\Delta t_p) - f(x,k,t)}{\Delta t_p} + \frac{\hbar k}{m} \frac{\mp f(x,k,t) \pm f(x \pm \Delta x, k, t)}{\Delta x}$$
$$+ \frac{C_0 U_1}{\hbar \Delta x \Delta k} \sum_{l=-1}^{1} a_{1,2}^l f(x, k + l\Delta k, t)$$
$$= \sum_{j=1}^{\infty} \left[ \frac{(-1)^j}{\hbar \Delta x \Delta k} C_{2j+1} U_{2j+1} \sum_{l=-j-1}^{j+1} a_{2j+1,2}^l f(x, k + l\Delta k, t) \right] \quad (12)$$

Here, we separate the local classical terms on the LHS and the nonlocal terms on the RHS. When uncertainty is large, the LHS becomes dominant, following classical dynamics (well-posed). As uncertainty approaches zero, the RHS grows rapidly, eventually leading to the condition $RHS \approx 0$ (ill-posed). In the latter case, corresponding to measurement, the LHS operates as a very small noise. We define the scenario where this noise is exactly zero as a "complete solution." In this case, the LHS perfectly satisfies the condition LHS = 0, which is precisely the classical Liouville equation. Thus, we postulate that under the constraints of uncertainty, the system adheres to quantum mechanics, which may appear as an illusion due to quasi-probability distributions. However, at the moment of measurement, such quasi-probability "illusions" vanish, and the system transitions to classical mechanics. This is objectively supported by earlier observations in Fig. 2d and Fig. 3c, which show that as uncertainty decreases, resonance diminishes, and the current level approaches zero.

Our results show that the real world closely resembles a numerical simulation[30]. In a world without uncertainty, there would be infinitely many solutions, making it impossible for our intellect to perceive them. Under perfect uncertainty—in other words, in our complete ignorance—all these possibilities reduce to a single solution, referred to as the "sacred timeline," as illustrated in Fig. 5f. This represents the main thread of spacetime as we experience it. Each time an act of measurement occurs by an observer, random boundary conditions are imposed on the timeline, generating new numerical solutions—that is, new timeline branches. As previously explained, these boundary conditions are random actions that fundamentally transcend spacetime, potentially resulting in phenomena that appear strange and discontinuous.

**Conclusion**

The Schrödinger equation inherently embodies the uncertainty principle, representing an intrinsic lack of complete knowledge about the system. Consequently, it provides only probabilistic predictions rather than definite measured states. In this work, we have discovered that in phase-space quantum mechanics, particularly through the Wigner-Moyal equation, uncertainty can be arbitrarily adjusted by tuning the observation window (i.e., the order of accuracy in our numerical scheme). As uncertainty decreases, the system transitions into a regime of quantum decoherence, where superpositions progressively collapse, ultimately reaching a measurement system with zero uncertainty, characterized by instantaneous nonlocal interactions—often referred to as "spooky action at a distance." Uncertainty, as its name implies, quantifies how well we understand a system, offering a profoundly clear and intuitive resolution to the long-standing quantum measurement problem. As uncertainty decreases, we acquire more precise information about the system, but simultaneously, spacetime is distorted. This suggests that all physical laws within spacetime hold only in the presence of sufficient uncertainty, which explains why we have been unable to uncover the mystery of measurements that transcend spacetime and physics laws[31,32]. Our results demonstrate that uncertainty tunability in quantum mechanics is the key to explaining observation-related phenomena in the real world, providing a foundational framework for this understanding.

**Competing interests**

The authors declare no competing interests.

**Data availability**

The data that support the findings of this study are available from the corresponding author upon reasonable request.

**Numerical Methods**

The finite difference coefficient $a^l_{j,m}$ for the $j$-th derivative with $m$-th order accuracy can be calculated by solving the linear system matrix:

$$\begin{pmatrix} 1 & 1 & \cdots & 1 & 1 \\ -p & -p+1 & \cdots & p-1 & p \\ (-p)^2 & (-p+1)^2 & \cdots & (p-1)^2 & p^2 \\ \cdots & \cdots & \cdots & \cdots & \cdots \\ \cdots & \cdots & \cdots & \cdots & \cdots \\ \cdots & \cdots & \cdots & \cdots & \cdots \\ (-p)^{2p} & (-p+1)^{2p} & \cdots & (p-1)^{2p} & p^{2p} \end{pmatrix} \begin{pmatrix} a^{-p}_{j,m} \\ a^{-p+1}_{j,m} \\ a^{-p+2}_{j,m} \\ \cdots \\ \cdots \\ \cdots \\ a^{p}_{j,m} \end{pmatrix} = \begin{pmatrix} 0 \\ 0 \\ 0 \\ \cdots \\ j! \\ \cdots \\ 0 \end{pmatrix} \quad (11)$$

where $p = \frac{j+m-1}{2}$ and the $j!$ is in the $(j+1)$-th low on the RHS. This method allows each derivative term to be easily discretized for any order and incorporated into the overall system matrix.

Equation (3) contains an infinite series of derivative terms, which must be truncated at a certain threshold. We retain only the matrix components on the RHS that are greater than $10^{-6}$ times the $j = 0$ term and discard the rest to construct an efficient sparse matrix.

In the resonant tunneling diode, the unwind injection boundary conditions are given by:

$$f(0,k) = \frac{m^* k_B T}{\pi \hbar^2} \ln\left[1 + \exp\left(-\frac{1}{k_B T}\left(\frac{\hbar^2 k^2}{2m^*} - u_f\right)\right)\right] \; for \; k > 0, \quad (12)$$

$$f(L,k) = \frac{m^* k_B T}{\pi \hbar^2} \ln\left[1 + \exp\left(-\frac{1}{k_B T}\left(\frac{\hbar^2 k^2}{2m^*} - u_f\right)\right)\right] \; for \; k < 0, \quad (13)$$

where $m^*$ is effective mass, $T$ is temperature, and $u_f$ is the Fermi level at the injection boundary. In this work, we used $m^* = 0.07 m_0$, $T = 77K$, and $u_f = 50mV$.

Electron density is obtained by numerically integrating the Wigner distribution function:

$$n(x) = \frac{1}{2\pi} \sum f(x,k) \Delta k. \quad (13)$$

Similarly, the current density can be expressed as:

$$J_{x+0.5\Delta x} = \frac{\Delta k}{2\pi}\left[\sum_{k<0}\frac{\hbar k}{m^*}f(x+\Delta x, k) + \sum_{k>0}\frac{\hbar k}{m^*}f(x,k)\right]. \quad (14)$$

To ensure current continuity, periodic boundary conditions are applied in k-space.

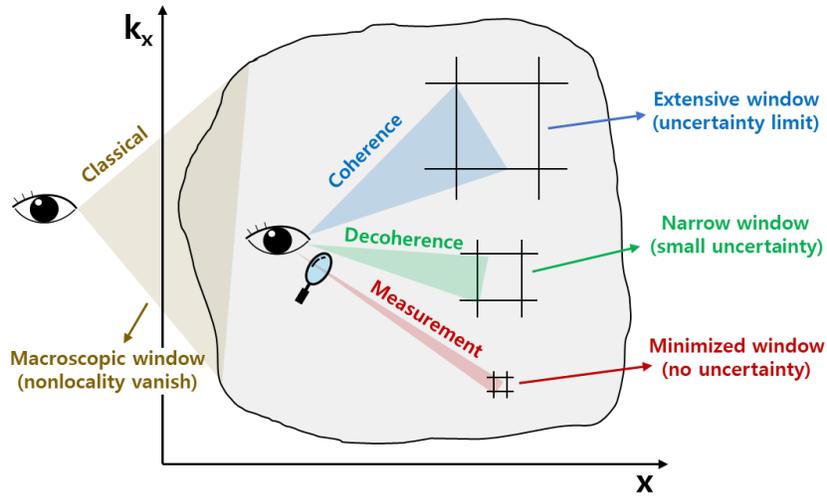

**Figure 1 – Illustration of the Relationship Between Observation Window, uncertainty, and System Behavior.**

This figure depicts how the size of the observation window and the level of uncertainty affect the behavior of a quantum system. As the observation window decreases, nonlocality increases, causing the numerical characteristics to transition through classical dynamics, coherent quantum dynamics, decoherence, and finally to a measurement regime without uncertainty.

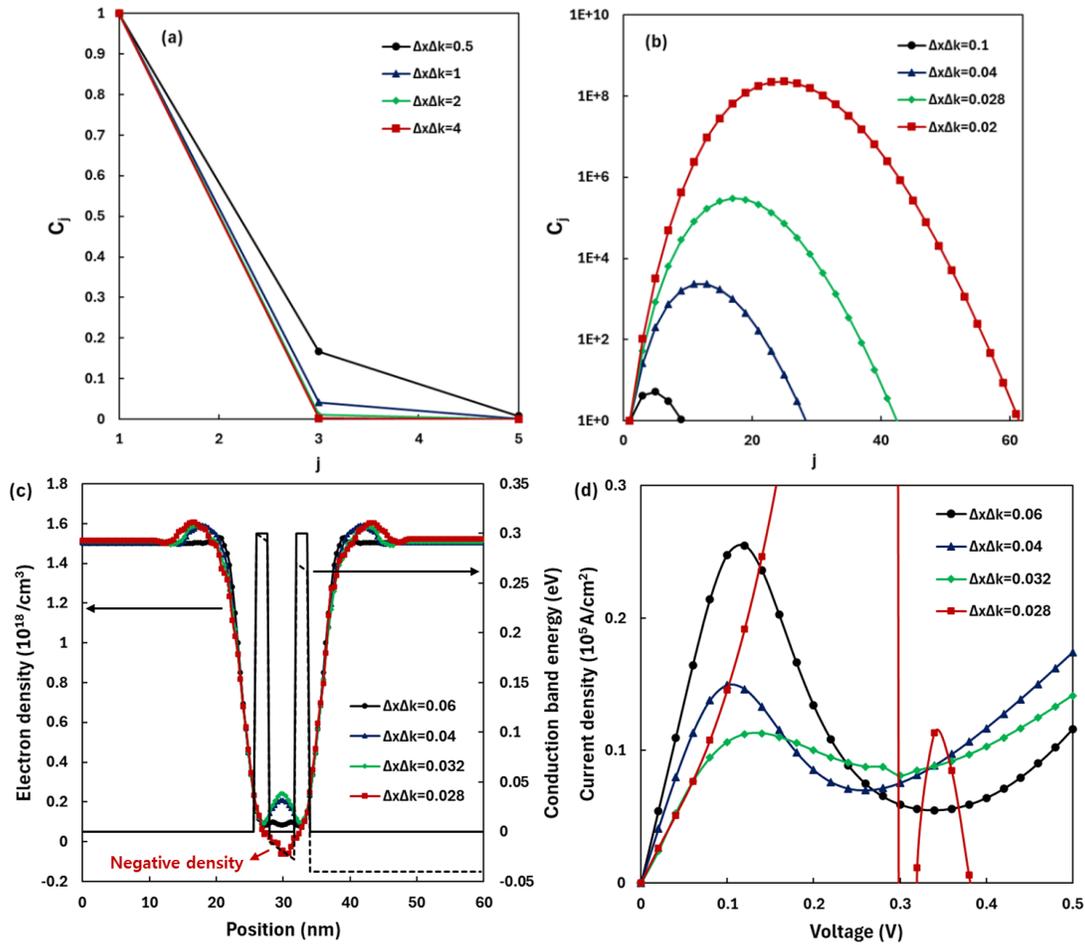

**Figure 2 – Numerical instability in cases of small mesh size.**

a. The nonlocal power $C_j$ in situations where $\Delta x \Delta k \geq 0.5$. As the mesh size increases, the high-order terms gradually vanish.
b. The nonlocal power $C_j$ in situations where $\Delta x \Delta k < 0.5$. As the mesh size decreases, the high-order terms increase rapidly.
c. Electron density and conduction band energy when zero bias is applied (equilibrium condition). When the mesh size decreases below a certain threshold, nonphysical results occur, such as the emergence of negative electron density in the quantum well region. Here, we fixed $\Delta x$ at 4 Å and adjusted $\Delta k$.
d. Mesh size-dependent current-voltage characteristics. At $\Delta x \Delta k = 0.028$, we observe that normal quantum behavior is completely disrupted. With even smaller mesh sizes, the situation worsens significantly.

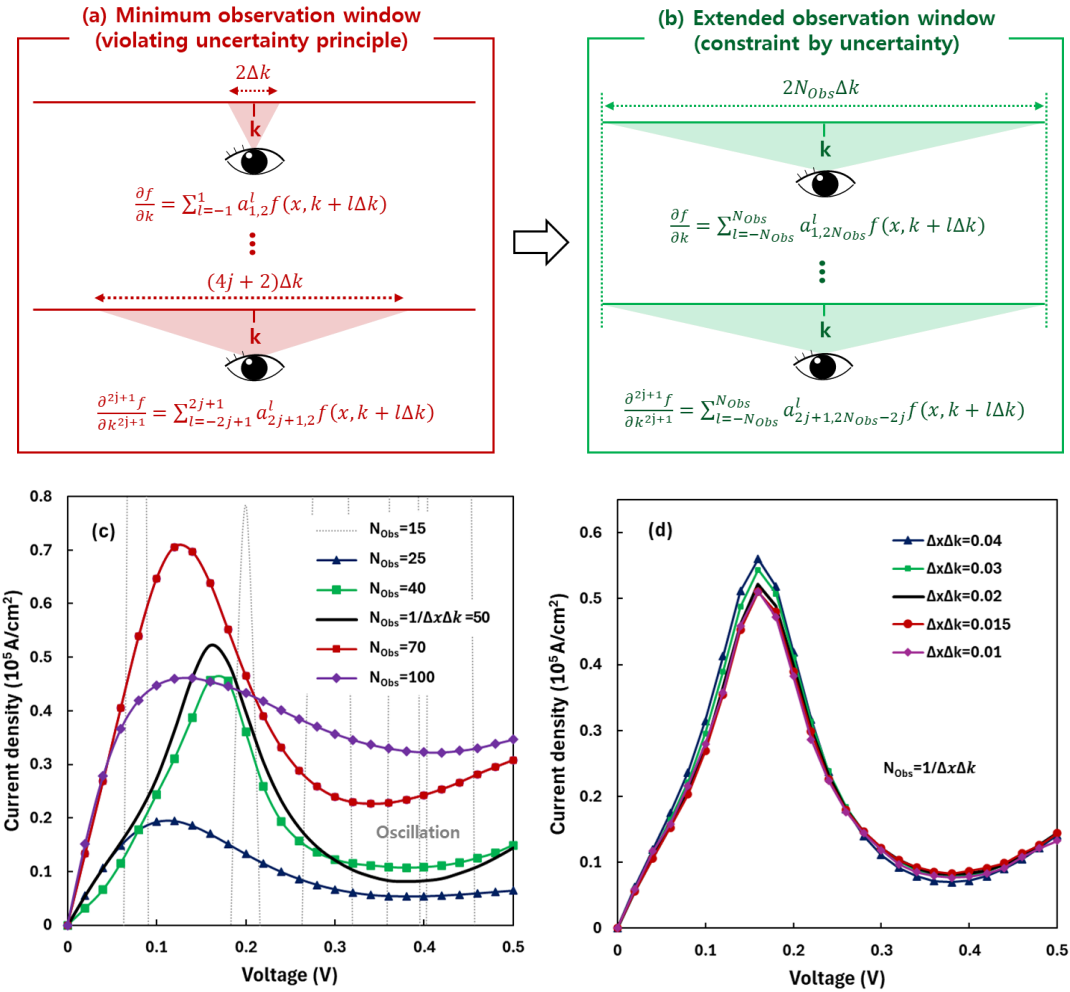

**Figure 3 – Considering the uncertainty principle's constraint by expanding the observation window.**

a. Observation window when using a second-order accuracy scheme. In this case, we observe data from an extremely narrow region, which violates the uncertainty principle.
b. To sufficiently widen the observation window within the constraints of the uncertainty principle, we need to observe data over a broader range when calculating partial derivative terms. This can be achieved by applying very high-order schemes to terms that, in low-order schemes, have observation windows smaller than a certain threshold.
c. Current-voltage characteristics according to the observation window at mesh spacing $\Delta x \Delta k = 0.02$. When the observation window is too small, the problem becomes ill-posed, exhibiting highly oscillatory and nonphysical result. By expanding the window, we progressively define a physical and well-posed problem. However, if the window becomes too large compared to the uncertainty limit, the quantum effects become faint.
d. Simulation results for various mesh spacings under the uncertainty limit condition ($N_{Obs} = N_{Uncertainty}$). Here, we fixed $\Delta x$ at 4 Å and adjusted $\Delta k$. Remarkably, the regularization of the equations considering the numerical uncertainty principle enables very stable simulations of the Moyal equation.

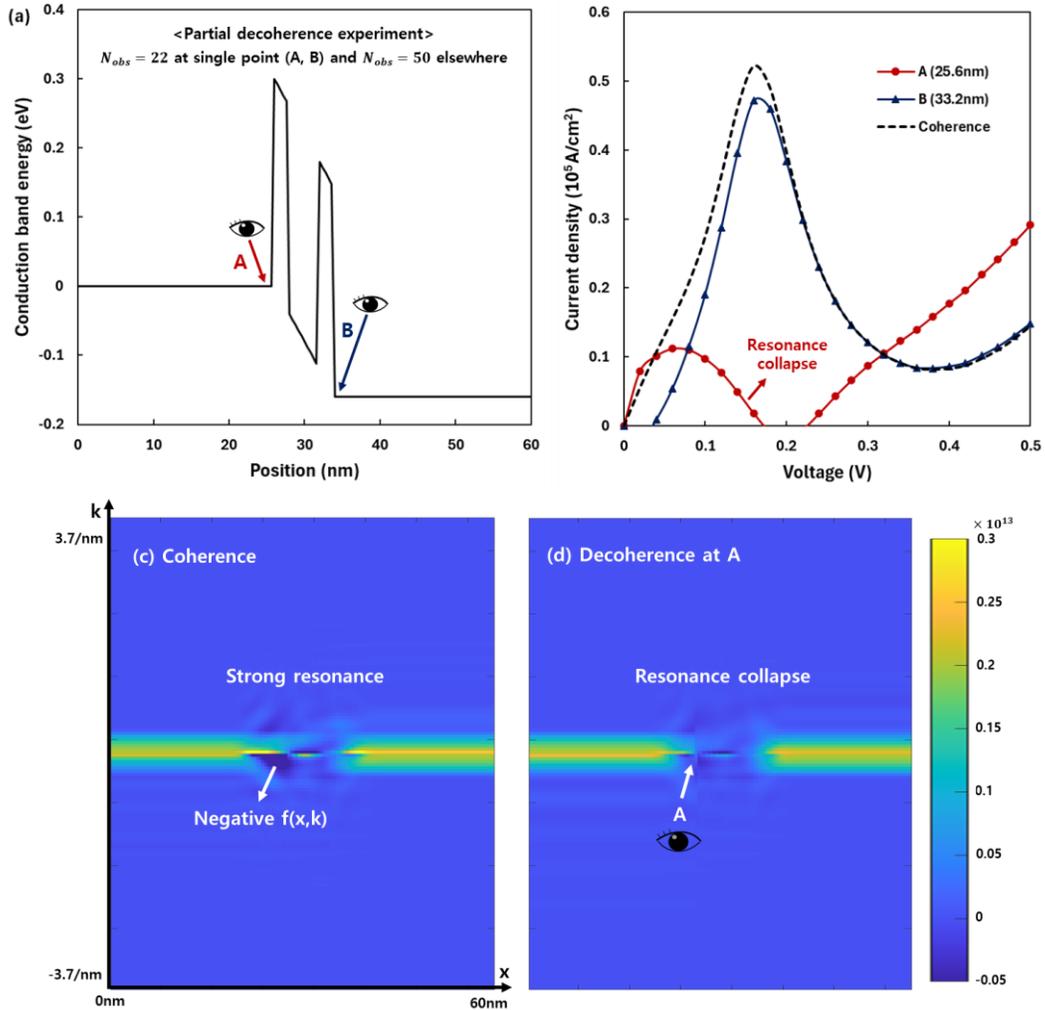

**Figure 4 – Partial decoherence experiment for resonant tunneling diode.**
  a. Decoherence induced by narrowing the observation window ($N_{Obs} = 22$) at specific points (A, B), while the rest of the system remained constrained by the uncertainty principle ($N_{Obs} = N_{Uncertainty}$). Here we used same mesh spacing as in Fig.3c.
  b. Current-voltage characteristics according to the position of decoherence. Decoherence induced just before traversing the double barrier (A) nearly eliminates resonance phenomena. Conversely, decoherence at point B shows similar result to the coherence case.
  c. Wigner distribution function in the coherence case ($V_{in} = 0.16V$, peak current). A prominent resonance pattern is observed.
  d. Wigner distribution function during decoherence at point A ($V_{in} = 0.16V$). In this scenario, the resonance patterns are barely observed.

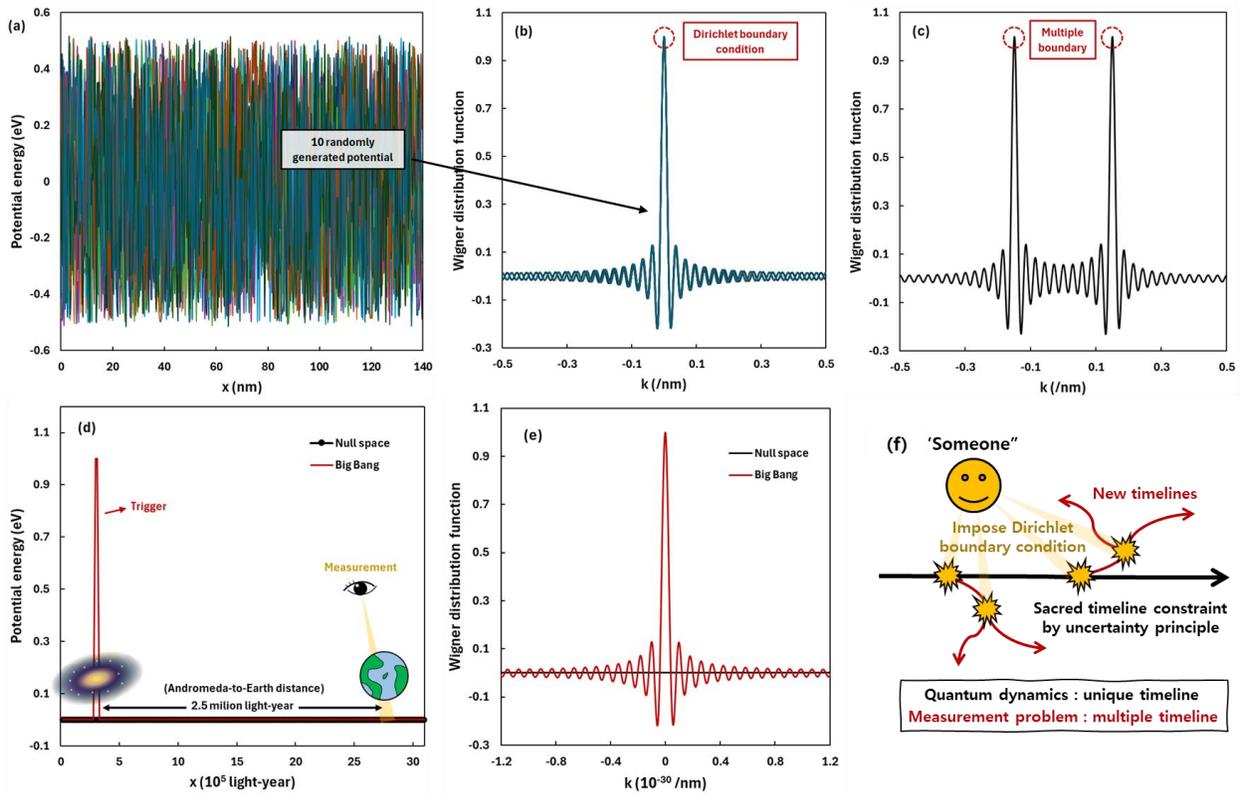

**Figure 5 – Measurement with minimized uncertainty.**

a. Ten randomly generated potential profiles. Potentials ranging from -0.5 to 0.5 eV were randomly assigned at each point.
b. Measurement results at x = 70 nm for ten random potential samples (solution of equation (9)). Although equation (9) depends on potential energy, consistent results are observed regardless of the potential profile.
c. Simulation with multiple boundary conditions. This results in multiple peak outcomes corresponding to each observation point.
d. A big bang toy model simulating the pre-universe null space and a pulse response triggered at a singularity. Measurements were replicated at points very far from the trigger point.
e. In the null space, potential nonlocality is absent, causing the nonlocal term to be zero and making measurement impossible. However, when a trigger occurs, measurements become possible even at distant locations, allowing, in principle, the measurement (creation) of matter across all space.
f. Solutions of quantum dynamic systems with uncertainty provide a unique and predictable "sacred timeline" governed by probability flows. However, measurement randomly imposes new Dirichlet boundary conditions within this unique timeline, creating numerous timeline branches.

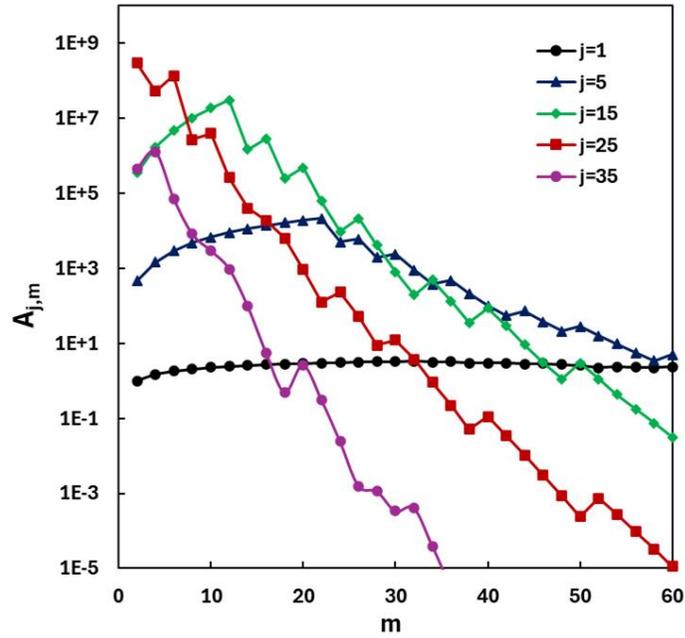

**Supplementary Material Figure 1 – The characteristics of $A_{j,m}$ for various $j$ and $m$.**
High-order differential terms (high $j$) exhibit a characteristic where $A_{j,m}$ decreases rapidly with increasing order of accuracy $m$.

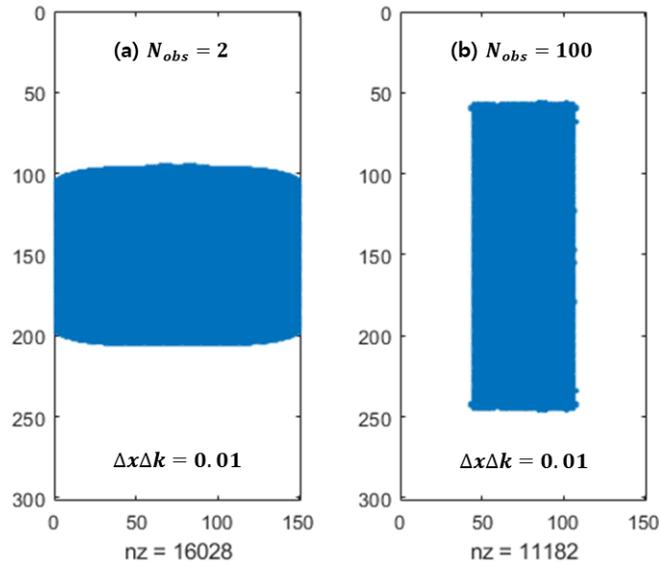

**Supplementary Material Figure 2 – The sparsity of the system matrix caused by nonlocal terms.**

The horizontal axis represents the mesh indices in x-space, while the vertical axis corresponds to the mesh indices in k-space. The $nz$ is the number of non-zero elements within the matrix. We considered a resonant tunneling diode, as shown in Fig. 2c.

(a) When the observation window is small, the uncertainty becomes minimal, and nonlocality increases sharply, causing the nonlocal components to spread widely across x-space and encroach upon the boundaries.

(b) Expanding the observation window increases sparsity and confines nonlocality to the quantum region near the double barrier. In areas where nonlocal terms are eliminated due to increased uncertainty, the system follows classical dynamics.